# The Controlled Large-Area Synthesis of Two Dimensional Metals

Tianyu Wang[1], Quanfeng He[1], Jingyang Zhang[1], Zhaoyi Ding[1], Fucheng Li[1], Yong Yang [1,2]*

[1.] Department of Mechanical Engineering, College of Engineering, City University of Hong Kong, Kowloon Tong, Kowloon, Hong Kong SAR, China

[2.] Department of Materials Science and Engineering, College of Engineering, City University of Hong Kong, Kowloon Tong, Kowloon, Hong Kong SAR, China

* Corresponding authors: Y. Y. (yongyang@cityu.edu.hk)

## Abstract

The rise of nanotechnology has been propelled by low dimensional metals. Albeit the long perceived importance, synthesis of freestanding metallic nanomembranes, or the so-called 2D metals, however has been restricted to simple metals with a very limited in-plane size (< 10 μm). In this work, we developed a low-cost method to synthesize 2D metals through polymer surface buckling enabled exfoliation. The 2D metals so obtained could be as chemically complex as high entropy alloys while possessing in-plane dimensions at the scale of bulk metals (> 1 cm). With our approach, we successfully synthesized a variety of 2D metals, such as 2D high entropy alloy and 2D metallic glass, with controllable geometries and morphologies. Moreover, our approach can be readily extended to non-metals and composites, thereby opening a large window to the fabrication of a wide range of 2D materials of technologic importance which have never been reported before.

**Introduction**

Freestanding two-dimensional (2D) materials, such as graphene and $MoS_2$, have been attracting tremendous interests over the past decade. Because of their unique nanostructures and reduced dimensionality[1], these 2D materials usually exhibit exceptional physical and chemical properties[2–5] and therefore have a great potential to be used in various advanced applications, such as highly sensitive sensors[6], flexible electronics[7,8], catalysts[9], and macromolecule separation systems[10,11]. In recent years, the family of 2D materials has been expanding, including not only single-layered graphene[12] and $MoS_2$[13] but also various ultrathin freestanding metallic films or the so-called 2D metals[14].

The atomic bonding in metals is intrinsically three dimensional. Therefore, metals were usually not considered as a typical 2D material. According to the recent first-principles calculations, single layered metals are thermodynamically stable only when their lateral size is less than 2 nm [15,16]. This poses a significant limitation on the size of 2D metals. In practice, it is extremely difficult to handle and deploy such tiny 2D metals for any real applications. Alternatively, one can increase the thickness of metals to overcome the issue of thermodynamic stability while keeping their surface-volume ratio high. The metals of such geometry are 2D in the sense that their physical and chemical properties are determined mainly by their surface. As seen in **Table 1**, the 2D metals reported so far are all based on a single element, having a thickness ($t$) ranging from 0.8 nm to 50 nm, a lateral dimension ($L$) from ~20 nm to ~8000 nm and an aspect ratio ($L/t$) from ~10 to ~$10^3$. Despite these geometric and chemical limitations, the obtained 2D metals displayed prominent and unusual properties that are impossible for their bulk counterparts[17,18]. For instance, 2D Au, Ag and Bi exhibit intense localized surface plasmon resonance (LSPR)[19,20], enabling the development of LSPR-based sensors[21]. In general, 2D metals possess good electrical conductivity, excellent flexibility, and highly active surfaces, making them a very competitive candidate material for advanced flexible electronics[22,23] and clean energy production[24,25].

To synthesize and fabricate 2D metals, people developed numerous top-down methods, such as mechanical compression[26] and nanolithography[27], as well as bottom-up methods, such as solution-based chemical synthesis[28–30]. Given the versatility and efficiency in handling pure metals, especially the noble ones[30–34], the solution based chemical synthesis method drew great attention and has been used to produce various 2D metals over the past decade. However, the shortcomings of the above methods are also evident. First, they were originally designed only for one or two specific pure metals and are difficult to be adapted to producing chemically and structurally complex alloys, such as 2D metallic glass and 2D high entropy alloys. Second, the in-plane size of the reported 2D metals was limited compared to that of graphene or $MoS_2$, which already reached the wafer scale according to the recent works[35,36]. In this work, we would like to develop a new method to address the above issues. Our method is versatile and capable of synthesizing 2D metals of extended chemical compositions. While their thickness ($t$) can be kept as thin as ~10 nm, their lateral dimension ($L$) can reach as large as ~10 mm with the corresponding aspect ratio ($L/t$) extended to $10^6$, which is about three orders of magnitude larger than those of the existing 2D metals [24,37–40].

Table 1. Comparison of the thickness and in-plane dimension of the 2D metals obtained via different methods

| Synthesis method | Metal Type | Thickness (t) (nm) | Lateral scale (L) (nm) | L/t | Ref. |
|---|---|---|---|---|---|
| Organic ligand-confined growth | Pd | 0.8 | 100-200 | 100-200 | [41] |
| | Pt | 2 | 500 | 250 | [33,42] |
| | | 0.8 | 50-500 | 50-500 | [43] |
| | Ag | 5-10 | 50-100 | 10 | [28] |
| | | 10-40 | 3000-8000 | 200-300 | [32] |
| | | 2 | 500 | 250 | [44] |
| | Cu | 50 | 500-1500 | 10-30 | [45] |
| | | 2 | 20 | ~10 | [29] |
| 2D template-confined growth | Au | 2.4 | 200-500 | 80-200 | [34] |
| Hydrothermal method | Rh | 0.4 | 500-600 | 1000-1500 | [46] |
| Nanoparticle assembly | Pt | 10 | 50-100 | 5-10 | [47] |
| Mechanical compression | Al / Ag | 2 | 7300 | ~3500 | [26] |
| Exfoliation method | Sb | 4 | 1000 | 250 | [25] |
| PSBEE (Present work) | Ti / HEA / MG | 10-50 | $10^4$-$10^7$ | $10^2$-$10^6$ | This work |

Note HEA represents high entropy alloy; MG represents metallic glass.

**Results**

**Fabrication of 2D metals**

Figure 1(a) illustrates the critical steps of the experimental method we developed to synthesize 2D metals. In our experiments, we selected the polyvinyl alcohol (PVA) hydrogel as the substrate material for the deposition of metallic thin films. As shown in Fig. 1(a), we prepared the precursory PVA substrate by 3D printing (see Methods). The as-printed PVA substrate exhibited surface striations along the printing direction (Fig. 1b-c). After that, we pressed the precursory PVA substrate against a single crystal Si (111) wafer with the root-mean-square surface roughness $R_{rms}$ = 0.5 nm at an elevated temperature of 120°C to 180°C for a period of 60 to 180 seconds for surface smoothening. As a result, the $R_{rms}$ of the hot embossed PVA could be reduced to ~ 1.6 nm (Fig. 1d-e). To facilitate handling, we attached the PVA substrate to a polyimide (PI) membrane during hot embossing. Subsequently, we deposited a thin metallic film with the desired chemical composition onto the PVA substrate via magnetron sputtering (Fig. 1f, see the details in Methods). Once film deposition was completed, we soaked the film-substrate system into pure water. Interestingly, with the swelling of the PVA substrate, the entire metallic film could exfoliate easily from the PVA substrate within minutes, floating in the water as a freestanding 2D material (please see Supplementary Video 1 for demonstration). In sharp contrast to all previous methods [3,21,46,48], we can easily fabricate 2D metals using our current method, which possesses a thickness at the scale of 10 nm and an exceedingly large in-plane dimension at the scale of 10 mm, as seen clearly in Figs. 1h-i and Supplementary Video 1.

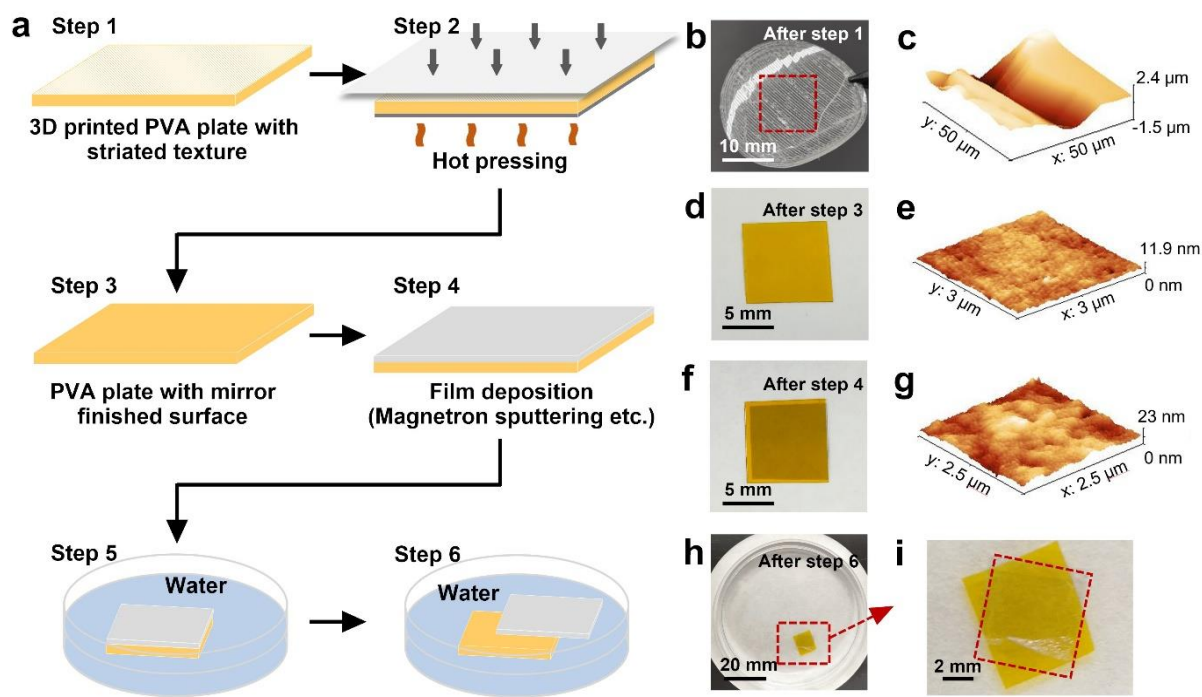

**Figure 1 Synthesis of 2D metals with our method of polymer surface buckling enabled exfoliation (PSBEE).** (**a**) Schematic illustration of the fabrication processes of 2D metals based on PSBEE, which involve the preparation of a Polyvinyl alcohol (PVA) substrate (step 1), thermal surface processing (step 2 -3), film deposition (step 4) and film exfoliation in water (step 5-6). (**b**) Photo of the as-printed PVA plate. (**c**) 3D surface topography of the as-printed PVA plate. (**d**) Photo of the thermally treated PVA substrate. (**e**) 3D surface topography of the thermally treated PVA substrate. (**f**) Photo of the Ti film deposited on the PVA substrate. (**g**) 3D surface topography of the Ti-PVA system. (**h-i**) Photos of the Ti film peeling off entirely from the PVA substrate.

Aside from the large area of the fabricated 2D metals, our method is applicable to a variety of chemical compositions, particularly the multi-component compositions which are not usually accessible by the traditional methods [5,14,41]. Figs 2 a-c show the scanning electron microscopy (SEM) images of three kinds of 2D metals fabricated using our method, i.e. 2D pure Titanium (Ti) (Fig. 2a), 2D FeCoNiCrNb high entropy alloy (HEA) with the nominal composition of FeCoNiCrNb$_{0.5}$ (in atomic percentage) (Fig. 2b), and 2D ZrCuAlNi metallic glass (MG) with the nominal composition of Zr$_{53}$Cu$_{29}$Al$_{12}$Ni$_6$ (Fig. 2c). Regardless of the chemical difference of these 2D metals (see Supplementary Figs 1-3 for the details of their chemical composition), their lateral dimension can easily reach the millimeter scale (Supplementary Fig. 4). To facilitate observation, these 2D metals were suspended across the circular hole of a copper grid with a diameter of ~60 µm. Interestingly, we could observe severe folding and some micro-cracking in these 2D metals; however, these damages could not spread over to cause overall breakage, hence implicative of good mechanical toughness that can afford the handling of the 2D metals during subsequent collection, transportation, and fixation. Figs 2 d-f show the high-resolution transmission electron microscopy (HRTEM) images of the 2D metals. Evidently, the 2D Ti is of a nanocrystalline structure (Fig. 2d), while the 2D HEA and 2D MG are of an amorphous structure (Figs. 2e-2f). These results are promising, which demonstrates that we could vary the atomic structure of the 2D metals by altering their chemical composition. To measure the thickness, we transferred the 2D metals from the Petri dish where they were fabricated to a Si substrate. As seen in the atomic force microscope (AFM) images (Figs. 2 g-i), it is clear that the detected thickness is ~21 nm for the 2D Ti (Fig. 2g), ~36 nm for the 2D HEA (Fig. 2h), and ~51nm for the 2D MG (Fig. 2i). In general, these thickness values are comparable to those of other 2D metals reported in the literature [28,32,45], being tabulated in Table 1. Notably, the thickness of the 2D metals fabricated via the current method can be easily reduced or increased by properly adjusting the processing parameters of magnetron sputtering, such as deposition time.

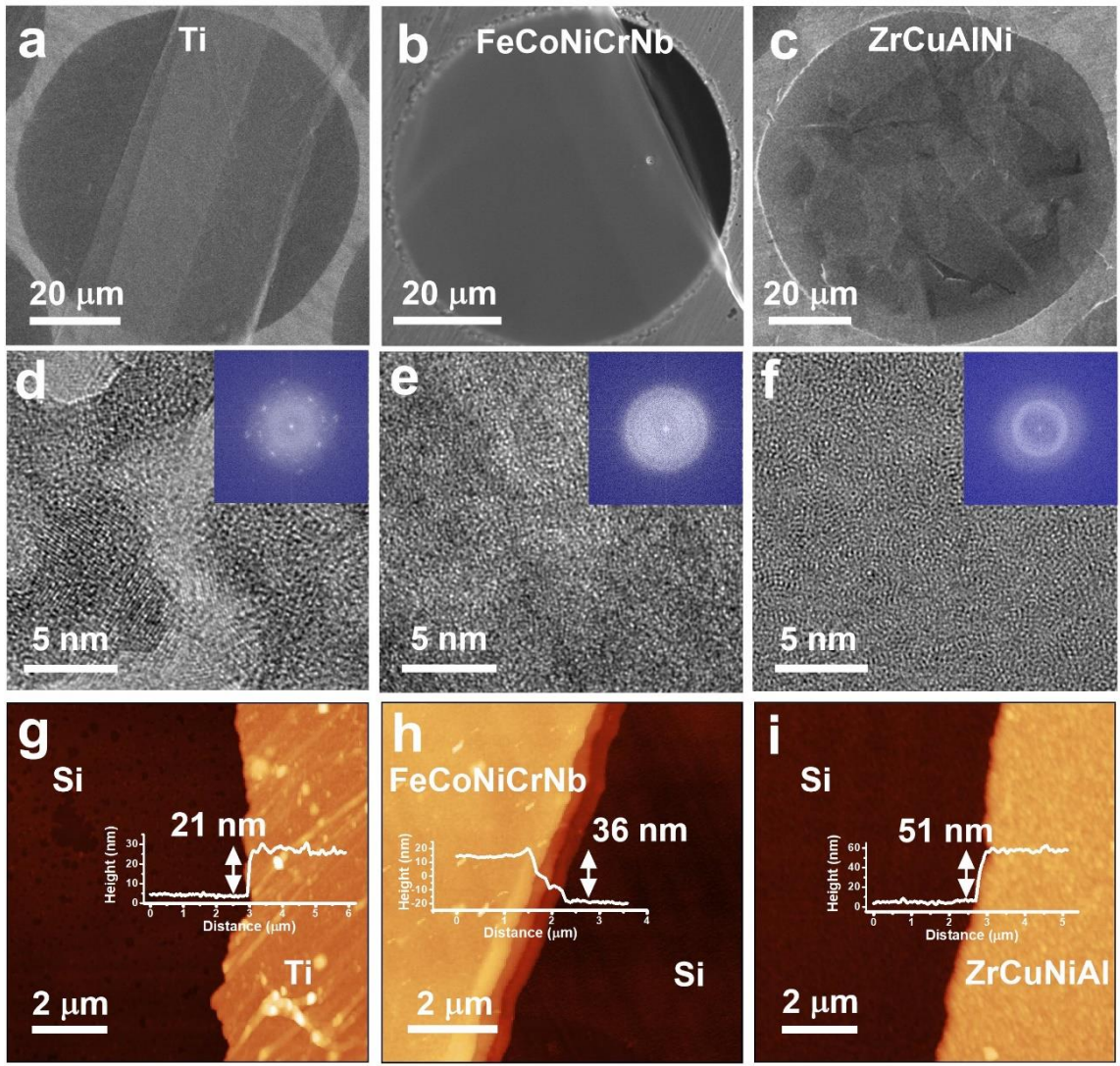

**Figure 2 Structural and geometric characterization of the as-fabricated 2D metals. (a-c)** SEM images of the 2D Ti, 2D FeCoNiCrNb and 2D ZrCuAlNi obtained via PSBEE, which were suspended over the copper grid for observation. **(d-f)** HRTEM images of the as-fabricated 2D Ti, 2D FeCoNiCrNb and 2D ZrCuAlNi. The insets show the corresponding fast fouier transformation (FFT) patterns of the HRTEM images. **(g-i)** AFM images of the 2D Ti, 2D FeCoNiCrNb and 2D ZrCuAlNi transferred to the Si substrate. The insets show the line scan across the edge of the 2D metals.

**Elastic characterization of the as-fabricated 2D metals**

After we obtained the 2D metals, we characterized their elastic moduli, which are relevant to the mechanics of film exfoliation as discussed in the later text, with nanoindentation modulus mapping (see Supplementary Figs.5a-c, and Methods for the experimental details). Interestingly, we found that the 2D Ti exhibited a slightly higher elastic modulus than thick Ti films (thickness > 1 μm) directly deposited on the silicon wafer (see Supplementary Fig. 5 d). Because $TiO_2$ possess a higher modulus than pure Ti according to the literature data[49,50], this behavior of stiffening, as observed on the 2D Ti, might be attributed to the formation of surface oxides. To confirm this, we carried out X-ray photoelectron spectroscopy (XPS) experiments on the 2D Ti. As seen in Supplementary Fig. 6, the presence of TiO2 is evident on the surface of the 2D Ti. Notably, when the thickness of the 2D Ti increases from ~20 nm to ~40 nm, its modulus reduces and approaches to that of the thick film, hence implicative of the diminishing effect of the surface oxides (Supplementary Fig. 5d). In contrast, the amorphous 2D FeCoNiCrNb and 2D ZrCuAlNi appeared softer than the corresponding thick films deposited on the silicon wafer. This behavior of softening agrees with the longstanding theory that the glass transition temperature and elastic modulus of amorphous films would reduce if the film thickness was decreased to the nanometer scale[51,52]. In other words, the effect of surface oxides on the 2D FeCoNiCrNb and 2D ZrCuAlNi, if there is any, is not as significant as that on the 2D Ti. As expected, one can clearly see that the surface of the obtained freestanding metal films plays a important role in their elastic properties. A more in-depth study is certainly needed to understand the mechanical behavior of these 2D metals, which however is already out of the scope of the current work and will be addressed soon.

**Theoretical modeling and finite element simulation of film exfoliation**
When water molecules diffuse into a PVA without physical constraints, the PVA swells under osmotic pressures. However, when a hard metal film was deposited onto the surface of the soft PVA, the free swelling would be constrained, which could cause the surface of the PVA to buckle[53–55], as illustrated in Figs. 3a-b. Interestingly, we observed that film exfoliation in our experiments was usually preceded by surface buckling after the film-substrate system was immersed in water (Fig. 3c, See Supplementary Video 2). According to Ref[56], the wavelength ($\lambda$) and amplitude ($\xi$) of the buckled surface (see the inset of Fig. 3d for the definition of $\lambda$ and $\xi$) being constrained by a hard film can be derived as $\lambda = \dfrac{2\pi h_f \left[ E_f / \left( 3E_s \right) \right]^{1/3}}{\left( 1 + \varepsilon_{cs} \right)\left( 1 + \zeta \right)^{1/3}}$ and

$\xi \approx \dfrac{h_f \sqrt{\varepsilon_{cs} / \varepsilon_c - 1}}{\sqrt{1 + \varepsilon_{cs}}\left( 1 + \zeta \right)^{1/3}}$, where $\zeta = 5\varepsilon_{cs}\left( 1 + \varepsilon_{cs} \right)/32$, $E_f$ and $E_s$ denote respectively the reduced elastic modulus of the film and substrate; $h_f$ represents the film thickness; and $\varepsilon_{cs}$ denotes the strain of the PVA substrate constrained by the film. To form buckling, $\varepsilon_{cs}$ has to reach the critical strain $\varepsilon_c = \dfrac{1}{4}\left( 3E_s / E_f \right)^{2/3}$ or above. Figs. 3e-f show the contour plots of the dimensionless variable $\lambda/h_f$ and $\xi/h_f$ as a function of $\varepsilon_{cs}$ and $E_f/E_s$ respectively. Evidently, the dimensionless amplitude ($\xi/h_f$) increases with the increasing $\varepsilon_{cs}$ and $E_f/E_s$; by comparison, the dimensionless wavelength $\lambda/h_f$ mainly depends on $E_f/E_s$ and rises with $E_f/E_s$. Note that the white region in Figs. 3e-f corresponds to $\varepsilon_{cs} < \varepsilon_c$ or the regime of no surface buckling. Physically, surface buckling induces an out-of-plane stress along the film-substrate interface. In our case, the out-of-plane stress should be tensile near the 'valley' of the buckled surface (see the inset of Fig. 3d) while compressive near the "peak." The maximum tensile out-of-plane stress, which occurs at exactly the bottom of the valley, can be expressed as[56]

$\sigma_z = \dfrac{E_f h_f^3}{12}\xi A^4 + E_f h_f \left( \dfrac{1}{4}\xi^2 A^2 - \dfrac{\varepsilon_{cs}}{1 + \varepsilon_{cs}} \right)\xi A^2$ where $A = \dfrac{\left( 1 + \zeta \right)^{1/3}}{h_f \left[ E_f / \left( 3E_s \right) \right]^{1/3}}$. As seen in Fig.

3g, similar to $\xi/h_f$, the dimensionless variable $\sigma_z/E_s$ increases with $\varepsilon_{cs}$ and $E_f/E_s$. In principle,

one can envision that a sufficiently high $\sigma_z$ would initiate film debonding and hence enables film exfoliation.

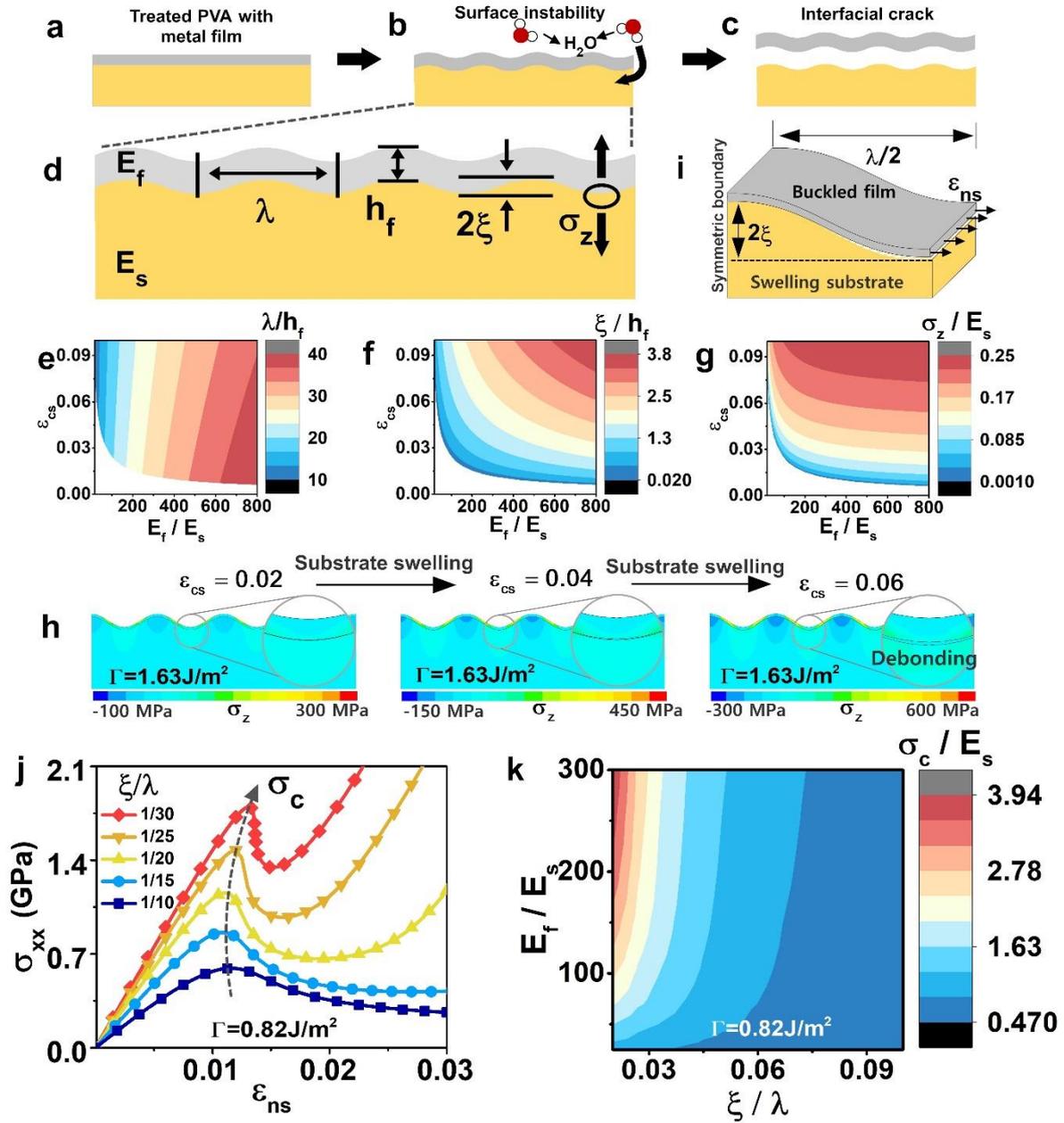

**Figure 3 Theoretical modeling and finite element simulation of surface buckling enabled film exfoliation.** Illustration of (**a**) the metallic film deposited on the PVA substrate, (**b**) water diffusion induced surface instability in the film-substrate system, and (**c**) complete film exfoliation subsequent to surface buckling. (**d**) The illustrated buckling geometry described by the wavelength ($\lambda$) and amplitude ($\xi$) of the surface oscillation together with the out-of-plane stress $\sigma_z$ acting at the film-substrate interface. Contour plots of **(e)** $\lambda/h_f$, **(f)** $\xi/h_f$ and **(g)** $\sigma_z/E_s$ as a function of $\varepsilon_{cs}$ and $E_f/E_s$. **(h)** Finite element simulations of the constrained swelling induced interfacial crack initiation at the buckled film-substrate interface. **(i)** Illustration of the geometry of the growing interface crack driven by continuously film stretching. **(j)** Plot of the critical film stress $\sigma_c$ versus the tensile strain $\varepsilon_{ns}$ in the film for various buckling geometries. **(k)** Contour plot of $\sigma_c/E_s$ as the function of $E_f/E_s$ and $\xi/\lambda$.

To gain quantitative insights into the process of film exfoliation, we carried out extensive finite element (FE) simulations by using a plane strain model with a cohesive-zone interfacial element (see Methods). To mimic the swelling of the PVA substrate constrained by the hard film, we assumed that the thermal expansion coefficient of the substrate was finite while that of the film was negligible. After that, we raised the overall temperature of the substrate-film system, which caused the substrate to expand in a similar fashion as swelling. However, because of the mismatch in the thermal expansion coefficient, the hard film impedes the substrate expansion. As a result, we can compute the constrained strain $\varepsilon_{cs}$ of the substrate based on the setting of the temperature change in our FE simulations (see Methods). Meanwhile, in order to initiate surface buckling, the substrate-film interface took on a sinusoidal profile with a very small amplitude, which mimics surface perturbation in real experiments. As seen in Supplementary Fig.7, the amplitude of the perturbation grows with the substrate expansion, indicative of surface buckling. As shown in Fig. 3h, the surface buckling results in stress localization, leading to an out-of-plane compressive stress around the 'peak' while an out-of-plane tensile stress around the 'valley.' This stress localization at the "valley" continues to grow with $\varepsilon_{cs,}$ which then causes excessive interfacial separation or debonding at a low $\varepsilon_{cs}$ value

if the interfacial toughness is small. By comparison, a tougher interface can tolerate a higher

$\varepsilon_{cs}$ value before debonding occurs (see Supplementary Fig.8). However, regardless of the

interfacial toughness, the general trend remains the same, which agrees with the theoretical modeling.

Once an interfacial crack is successfully initiated at the bottom of the 'valley', it can grow further if there is a tensile stress in the metallic film[57]. Assuming that the edge of the film can be stretched continuously at the strain $\varepsilon_{ns}$, as depicted in Fig. 3i, we can study the growth of the

interfacial crack for different geometries of surface buckling. As shown in Fig. 3j, when $\varepsilon_{ns}$

increases, the film stress $\sigma_{xx}$ rises initially but drops at a certain point when $\sigma_{xx}$ reaches a critical value $\sigma_c$. The sudden drop of the film stress marks the onset of the interfacial crack growth.

Therefore, we can take $\sigma_c$ as the indicator for the easiness or difficulty for the interfacial crack

growth. Depending on the geometry of the buckled surface, the film stress either continues to

decline or goes up again with further crack growth. In theory, the interfacial crack growth is

driven by the tensile stress in the film, which can be developed during film deposition and

substrate swelling. According to the dimensional analysis[58], we can easily derive that $\sigma_c/E_s$

$=f(\xi/\lambda, E_f/E_s)$ for a given $\Gamma$ value, where $f$ is a continuous function that can be determined by

the FE simulations. As shown in Fig. 3k, the $\sigma_c/E_s$ value is determined mainly by the ratio of

$\xi/\lambda$ when the $E_f/E_s$ value is large. The general trend is that the larger is the $\xi/\lambda$ ratio the smaller

is $\sigma_c$. In other words, the rougher is the buckled surface the easier is interfacial cracking. Here,

it is worth noting that the $\sigma_c$ value decreases with the decreasing $E_f/E_s$ if $\xi/\lambda$ is very small, as manifested by the region on the left bottom corner of Fig. 3k. On the other hand, as seen in Supplementary Fig. 9, the $\xi/\lambda$ ratio also scales with $E_f/E_s$ and $\varepsilon_{cs}$ in a similar manner as $\sigma_z/E_s$. Therefore, the $E_f/E_s$ ratio and $\varepsilon_{cs}$ determine not only interfacial crack nucleation but also interfacial crack propagation. In principle, one needs to raise $\varepsilon_{cs}$ for the nucleation and propagation of an interfacial crack; however, the effect of $E_f/E_s$ is twofolds: it needs to be increased for surface roughening (Supplementary Fig. 9) and the nucleation of the interfacial crack (Fig. 3g) but needs to be decreased for its subsequent propagation (Fig. 3k). This is a rather complicated process which requires one to carefully tune $E_f/E_s$ and $\varepsilon_{cs}$ in such a way that film exfoliation in real experiments can be facilitated.

**Discussion**

**Film stress and metal-hydrogel interfacial toughness**

Aside from the elastic mismatch, a tensile stress in the film is required to drive an interfacial crack to propagate after its nucleation. Because of the amorphous nature of our hydrogel substrate and the low substrate temperature during film deposition, the common mechanisms for the development of residual stress, such as lattice mismatch and atom diffusion into grain boundaries [59,60], are unlikely to apply. Therefore, the coalescence of atom "islands" during film deposition, as discussed in the literature[61–63], offers the possible mechanism for the development of tensile residual stress in our thin metallic films. As seen in Supplementary Video 1, the substrate-film system clearly bent towards the side of the metallic film right after the film deposition, hence indicative of tensile film stress. In addition to the tensile film stress, the relative low toughness of the metal-hydrogel interface also facilitates the process of film exfoliation. According to Ref.[64], the interfacial toughness between hydrogel and metal could be very weak, particularly so in a liquid environment, such as water. Physically, it was already shown via XPS in the early works[65–67] that the atomic/molecular bonding between hydrogel and metal is mainly through surface oxides and metallic atoms. However, the diffusion of water molecules into the interface can easily undermine such weak bonds with the assistance of flexible C-C bonds[68]. Therefore, the successful film exfoliation from the PVA substrate immersed in water, as witnessed in our experiments, reflects a rather weak hydrogel-metal interface, which is consistent with the prior results reported in the literature[64].

**Transition of the morphology of 2D metals: From sheets to scrolls**

As aforementioned, film exfoliation in our experiments depends on the ratio of $E_f/E_s$ and the constrained strain $\varepsilon_{cs}$. Here, we would like to demonstrate that one may change the morphology of the obtained 2D metals by tuning these two design parameters. To this end, we prepared solution based PVA (S-PVA) as an alternative substrate for the deposition of metal films (see Methods). We measured the modulus of the S-PVA and the thermally pressed PVA (T-PVA) (Fig. 1) using dynamic mechanical analyses (DMA) (see Methods). Compared to T-PVA, which has an elastic modulus of ~2 GPa at room temperature, S-PVA has a much lower elastic modulus (~0.1 GPa) because of its higher concentration of water molecules. To understand the swelling behavior of these two different substrates, we soaked them into water and measured their in-plane strain with digital image correlation (DIC) analyses (see Methods). As seen in Supplementary Fig. 10, our DIC results clearly show that, under the same experimental condition, S-PVA undergoes free swelling to a lesser degree than T-PVA, which suggests a lesser degree of constraint or a smaller strain $\varepsilon_{cs}$ in the hydrogel-metal system. In such a case, one could envision that the mechanics of surface buckling would lead to a different film exfoliation behavior if one replaces T-PVA with S-PVA for the synthesis of 2D metals. In general, the S-PVA substrate tends to yield a larger $E_f/E_s$ ratio but smaller $\varepsilon_{cs}$ than the T-PVA substrate. Consequently, surface buckling with a large $\xi/\lambda$ ratio becomes relatively difficult on the S-PVA, therefore defying interfacial crack nucleation. In such a case, film exfoliation, if there is any, tends to take place from existing cracks or sites of geometric discontinuity, such as the edges of a metallic film. Consequently, the 2D metals tend to roll up into scrolls for the release of the residual stress [23] (see Supplementary Video 3) rather than peel off as a sheet from the hydrogel substrate (see Fig. 4a). Fig. 4b shows the nanoscrolls obtained by applying the S-PVA to the FeCoNiCrNb-hydrogel system. Here, it may be worth mentioning that, compared to the freestanding FeCoNiCrNb nanosheet (Fig. 4c), the FeCoNiCrNb nanoscrolls produced by rolling-up delamination could be more useful in various energy-related applications, such as electrocatalysis, because of their curved morphology.

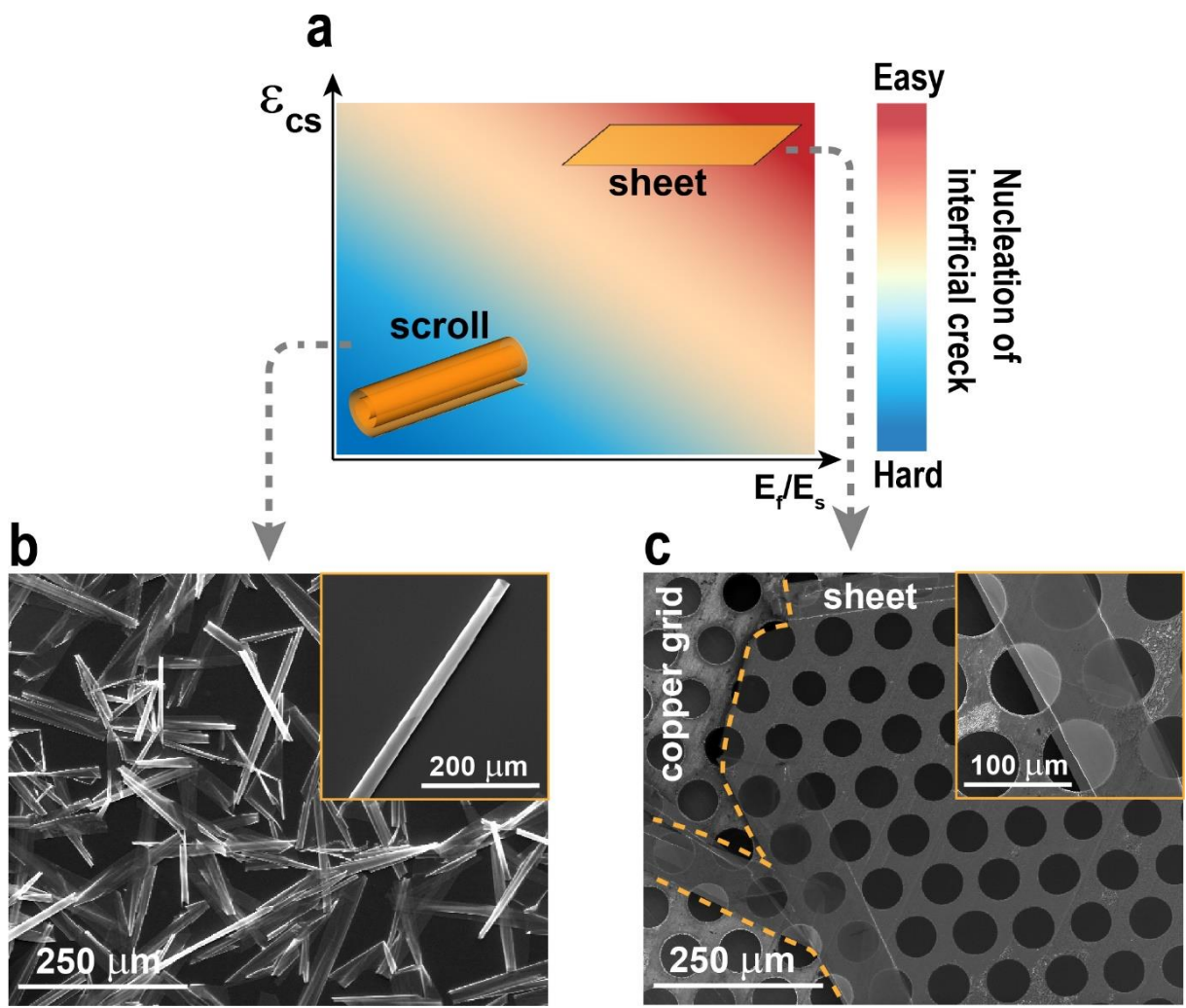

**Figure 4 Transition of the morphology of the 2D FeCoNiCrNb from nanosheets to nanoscrolls.** (**a**) Illustration for the transition from sheets to scrolls as controlled by $\varepsilon_{cs}$ and $E_f/E_s$. Note that the color represents the easiness of interfacial crack initiation. (**b**) SEM images of the FeCoNiCrNb nanoscrolls on Si. (**c**) SEM images of the FeCoNiCrNb free-standing nanosheets on a copper grid.

## Control of the geometry of 2D metals

Apart from the diversity of atomic structure, chemistry and morphology of the obtained 2D metals, the other advantage of our current method is that we are now able to fabricate 2D metals with a controlled shape in mass production. From the application viewpoint, this is of great importance to biomedical engineering because the shape of 2D metals can affect how these metallic nanosheets enter and interfere with human cells[69,70]. To synthesize 2D metals with a controlled shape, we can cover the PVA surface (see Fig. 1d) with different masks (see Fig. 5 a-e) during film deposition. As a result, the deposited metallic film on the patterned PVA surface can exfoliate and form 2D metals of the selected shape, as exemplified by Fig. 5 f-h. Here, it should be stressed that the final shape of the 2D metals depends not only on the mask but also on the thickness and physical properties of the material. Our current results suggest that the underlying physics of the shape control for our obtained 2D metals may be related to the mechanics of folding and self-organization, which we will address in the near future.

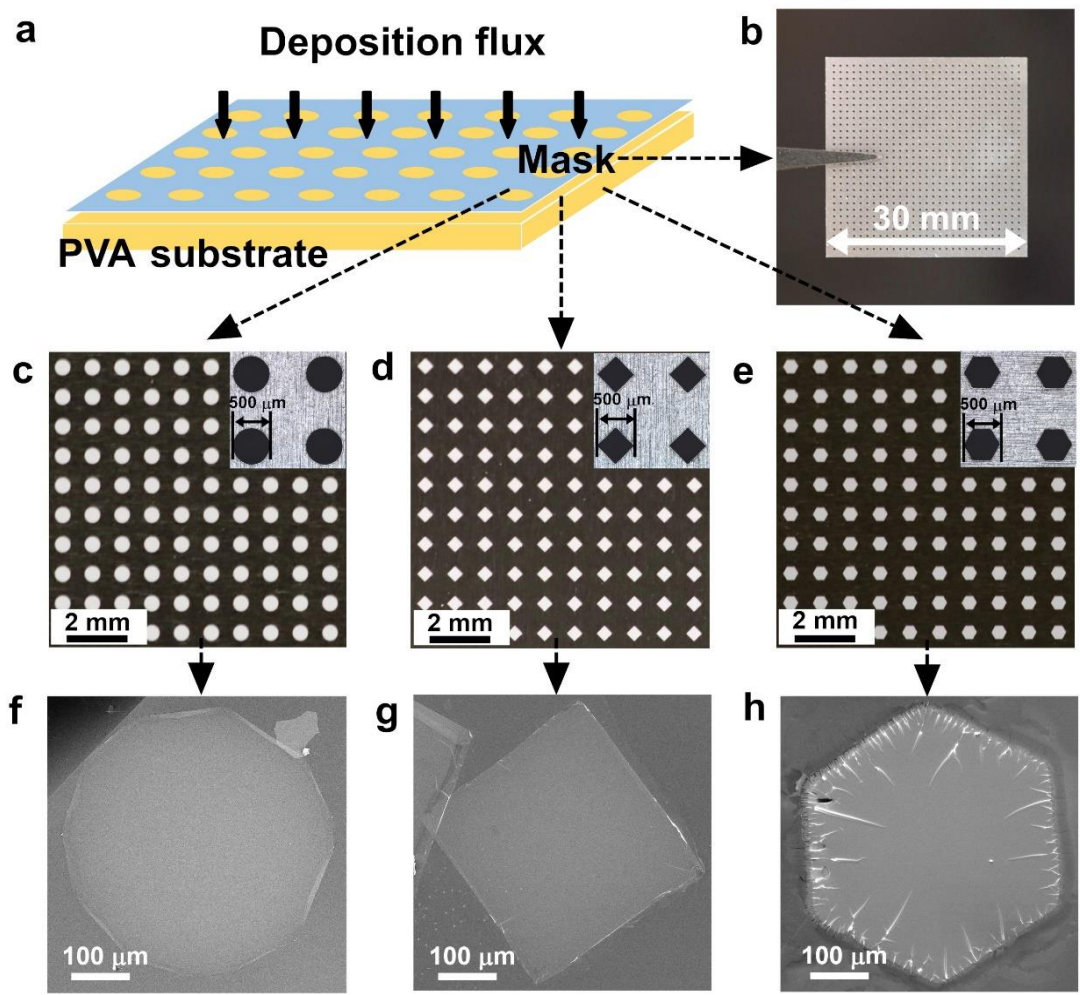

**Figure 5 Controlling the geometry of the fabricated 2D metals. (a)** Illustration of the deposition process with a mask. (**b**) Low magnification photo of the deposition mask. Photos of the deposition masks with (**c**) round, (**d**) square and (**e**) hexagon holes. The insets show the high magnification images of the masks. (**f-h**) SEM images of the round, square and hexagon shaped 2D FeCoNiCrNb fabricated via the masked deposition.

**Beyond the synthesis of 2D metals**

Although the current work is focused on 2D metals, however, our method is applicable to other materials owing to the versatility of the state-of-the-art deposition techniques. Very interestingly, the outcome of our research already demonstrated that, in addition to the large-sized 2D metals, one can also obtain large-sized non-metallic nanosheets, such as the free-standing amorphous silicon nanosheets (Supplementary Figs 11a-c) and even large-sized composite nanosheets, such as the freestanding Ag-C composite nanosheet (Supplementary Figs 11b-d). On one hand, synthesis of these freestanding nanosheets, either metallic or non-metallic, greatly enrich the family of 2D materials, which opens up a large window towards the discovery of new 2D materials with technologic significance; on the other hand, we also believe that the chemical, structural, morphologic and geometric diversities brought about by our method, which is not seen in the prior works, also give rise to new opportunities for a variety of applications, which could be related to bioengineering and nanotechnology, such as nano-medicine[69] and nano-sensing[71].

**Methods**

**3D printing of precursory PVA**

The precursory PVA substrate was prepared using the commercial Ultimaker® 3D printer with PVA filaments. The extruder (AA0.4 type) was heated to about 500 K before a single layer of PVA was printed at the speed of 20 mm/s on a cold glass working plate at room temperature. After printing, the precursory PVA was removed from the working plate for hot pressing.

**Magnetron sputtering**

The direct current (DC) magnetron sputtering system was used to deposit metal films on the treated PVA surface. The sputtering targets had a high chemical purity and custom made by the local companies. Before film deposition, we applied a 1min plasma etching to the target to remove oxides or contamination. The deposition process of the three targets (Ti, ZrCuAlNi, and FeCoNiCrNb) all worked at 100W for 4min with Ar (99.9% purity) as the working gas at an overall pressure of 10 mTorr. During the deposition process, the target-substrate distance was kept at around 15cm with the substrate rotating at 10 rpm.

**Nanoindentation modulus mapping**

We performed the modulus mapping tests using the TI-950 Hystron nanoindentation platform under the contact mode. The transducer is equipped with a Berkovich diamond tip with a tip radius of ~180nm. The 2D metals (Ti, ZrCuAlNi, and FeCoNiCrNb) were placed on the silicon wafer to facilitate the measurement. During the modulus mapping, the tip was scanned over an area of 10×10 μm with the oscillation frequency of 200 Hz. The set-point force and load amplitude applied were 5-7 μN and 2-3 μN, respectively, depending on the property of the 2D metals. The stiffness, contact force and oscillation amplitude were directly measured from the sample response. Based on the estimated tip radius, we can easily calculate the elastic modulus of the 2D metals. In the meantime, we deposited relatively thick films (>1 μm) on the Si wafer and measured their elastic modulus as the benchmark.

**Finite element (FE) analysis**

The FE simulation was carried out using the commercial package ANSYS®. The 8-node plane element Plane183 was used to mesh the model. The thermal expansion coefficient of the substrate was set to $2.5\times10^{-3}$ $K^{-1}$ while that of the film was set to $1\times10^{-10}$ $K^{-1}$, which was negligibly small. Therefore, we could mimic the solvent-induced substrate swelling by assigning uniform temperature increment to the FE model. To simulate interfacial debonding, cohesive zone material (CZM) model was adopted in the mesh of the film-substrate interface. During the simulations, we tuned the interfacial toughness by adjusting $\sigma_{max}$, the maximum normal traction at the interface, and $\bar{\delta}_n$, the normal separation across the interface where the maximum normal traction is attained. The interfacial toughness could be easily calculated as $\Gamma = e\,\sigma_{max}\bar{\delta}_n$, where e is the Euler's number.

**Preparation of solution based PVA (S-PVA)**

The solution based PVA (0.02g/ml) was prepared by dissolving PVA filaments into deionized water, which was subsequently spread evenly onto a polyimide support. After curing for about 1 hour at 360 K in the drying oven, the solution formed a layer of PVA membrane on the polyimide surface. After that, the membrane was transferred into a vacuum chamber for storage before usage.

**Dynamic mechanical analyses (DMA)**

We used the METTLER® TOLEDO DMA system to measure the elastic modulus of the T-PVA and the S-PVA. The PVA samples with controlled geometry were fixed on the titanium clamps with the effective length of 15mm and tested under tension mode. The stretching force oscillated around the setpoint of 1.125N with the amplitude of 0.75N at a frequency of 1Hz.

The temperature of the chamber increased from room temperature (298K) to 430K at 2K/s during the test. The modulus of the T-PVA and the S-PVA were calculated using the storage modulus and loss modulus obtained from the tests.

**Digital image correlation**

To carry out the digital image correlation, we first introduced the tracking speckles onto the PVA surface. This was done by spraying 0.5 μm diamond particles onto a flat clean glass plate. After that, the glass plate was placed into a drying cabinet for about 10 minutes to let the solvent evaporate thoroughly. After that, the diamond particles was transferred onto the PVA surface as tracking speckles through mechanical contact. Next, we immersed the PVA into deionized water and recorded the swelling processes using an optical microscope (Leica DM 2700M). The digital image correlation analysis was carried out on the commercial Vic-2D software and the picture at the onset of swelling was selected as the reference.

**Acknowledgments**

The research of YY is supported by the Research Grant Council (RGC) through the General Research Fund (GRF) with the grant number CityU11213118 and CityU11209317. Part of the research is supported by City University of Hong Kong with the project number 9610391.

**The authors declare no conflict of financial interest.**

**Contribution of authors**

YY supervised the project; TW and YY conceived the idea; TW developed the method and performed the finite element simulations; QH carried out the electron microscopy tests and image correlation analyses; JZ performed the nanoindentation modulus mapping; DZ developed the experimental set-up for 3D printing and carried out the XPS analyses; FL designed the experiments for the synthesis of 2D metals with engineered shape and size. TW and YY wrote the manuscript. All authors all contributed to the analysis and interpretation of the data and have opportunities to comment on the final version of the manuscript.

**Supplementary Information**

**The Controlled Large-Area Synthesis of Two Dimensional Metals**

Tianyu Wang[1], Quanfeng He[1], Jingyang Zhang[1], Zhaoyi Ding[1], Fucheng Li[1], Yong Yang [1,2]*

1. Department of Mechanical Engineering, College of Engineering, City University of Hong

Kong, Kowloon Tong, Kowloon, Hong Kong SAR, China

2. Department of Materials Science and Engineering, College of Engineering, City University

of Hong Kong, Kowloon Tong, Kowloon, Hong Kong SAR, China

* Corresponding authors: Y. Y. (yongyang@cityu.edu.hk)

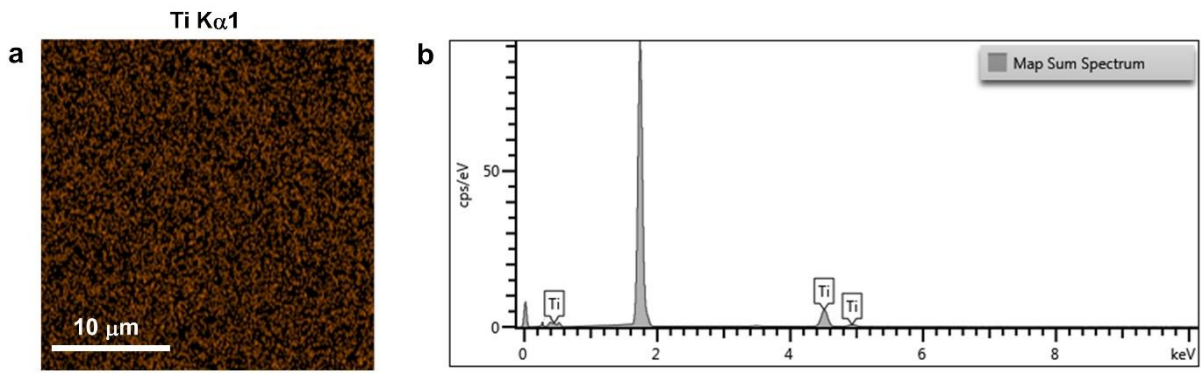

**Supplementary Figure 1** (a) The composition mapping and (b) corresponding energy spectrum of the 2D Ti.

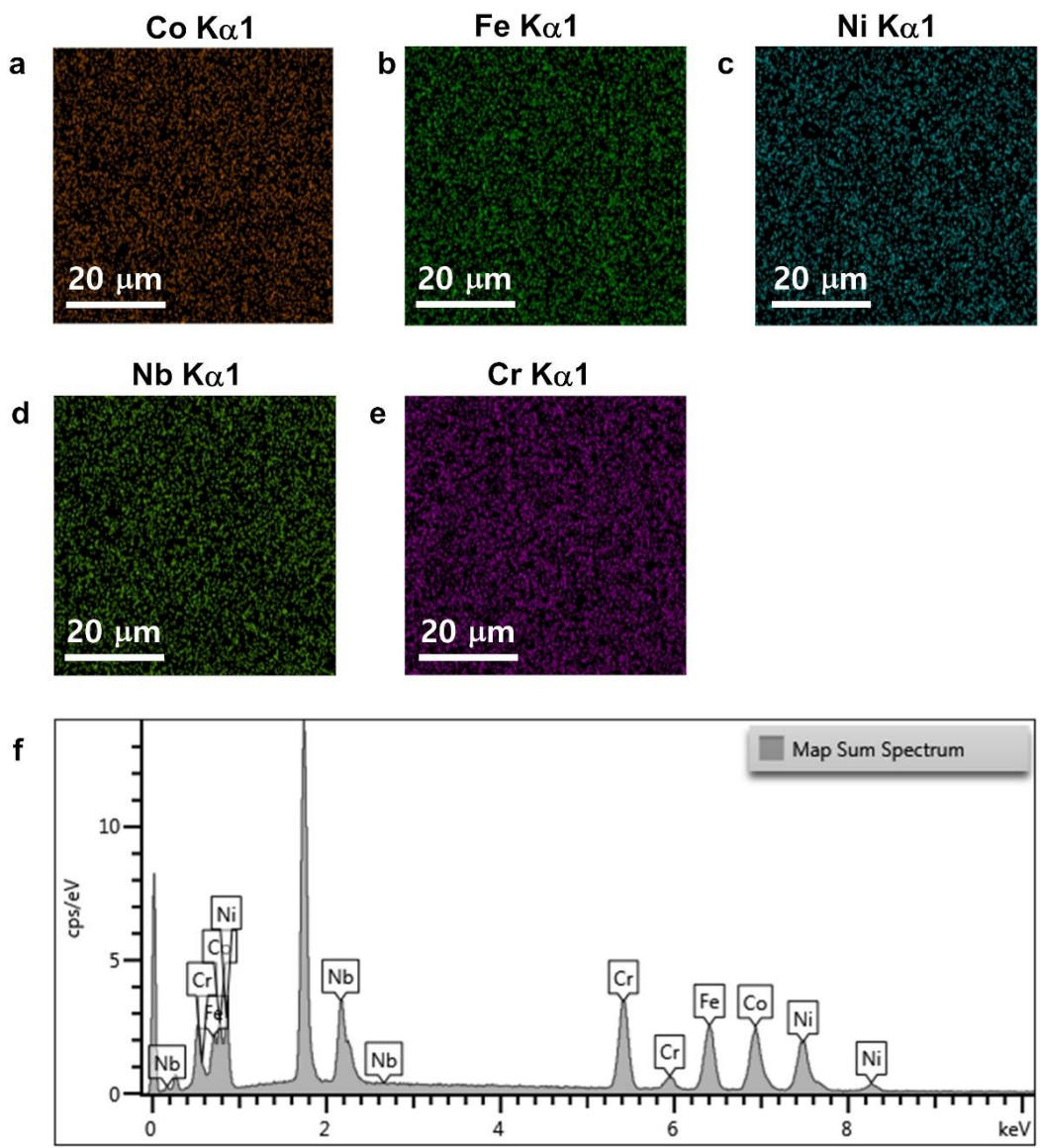

**Supplementary Figure 2** (a)-(e) The composition mapping and (b) corresponding energy spectrum of the 2D FeCoNiCrNb.

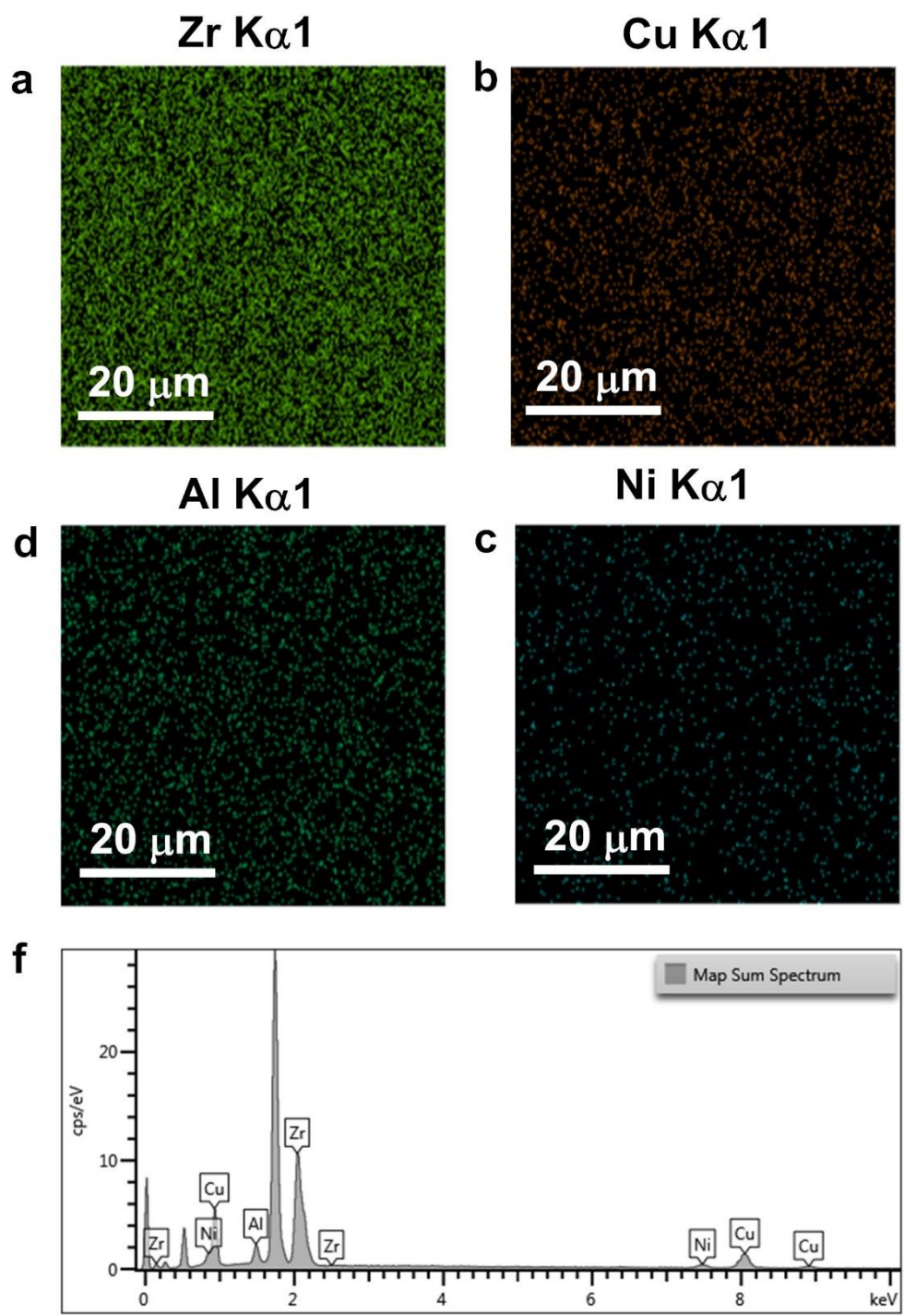

**Supplementary Figure 3** (a)-(e) The composition mapping and (b) corresponding energy spectrum of the 2D ZrCuAlNi.

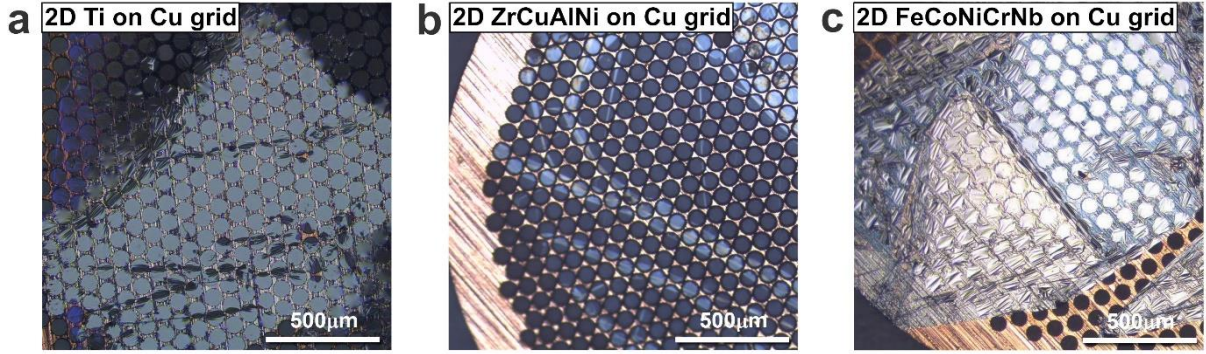

**Supplementary Figure 4** The optical images of the as-fabricated (a) 2D Ti, (b) 2D and (c) 2D FeCoNiCrNb transferred onto the copper grid in order to facilitate experimental observation.

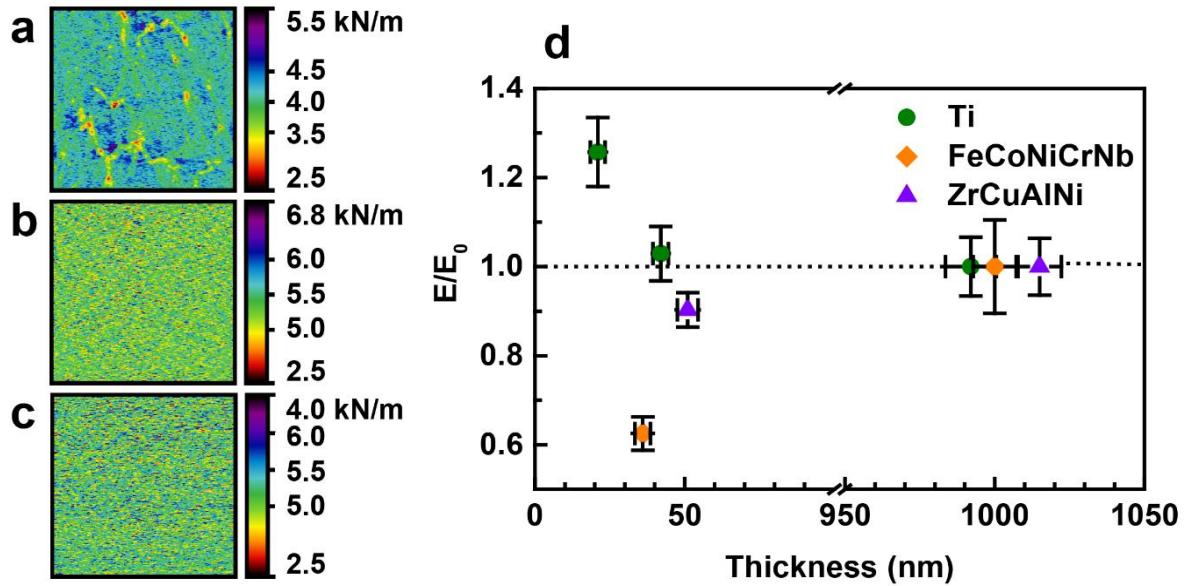

**Supplementary Figure 5 The elastic modulus of the 2D metals.** The stiffness contour plots of (**a**) the 2D Ti with a thickness of 21nm, (**b**) the 2D FeCoNiCrNb with a thickness of 36nm, and (**c**) the 2D ZrCuNiAl with a thickness of 51nm. (**a**) the thickness dependency of the elastic modulus measured for different 2D metals, being normalized by the modulus ($E_0$) of the thick film deposited on the Si substrate.Note that $E_0 = 74.1\pm4.9$ GPa for Ti, $215.5\pm22.6$ GPa for FeCoNiCrNb, and $93.1\pm5.7$ GPa for ZrCuAlNi. The error bars come from multiple experiments.

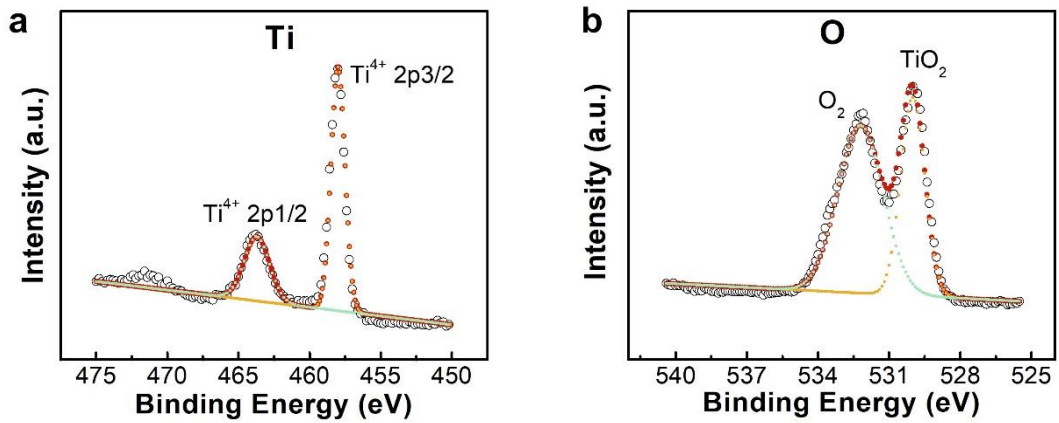

**Supplementary Figure 6 The XPS spectrum obtained from the surface of the as-fabricated 2D Ti nanosheet.** (**a**) the signal intensity distribution of the Ti element. (**b**) the signal intensity distribution of the O element.

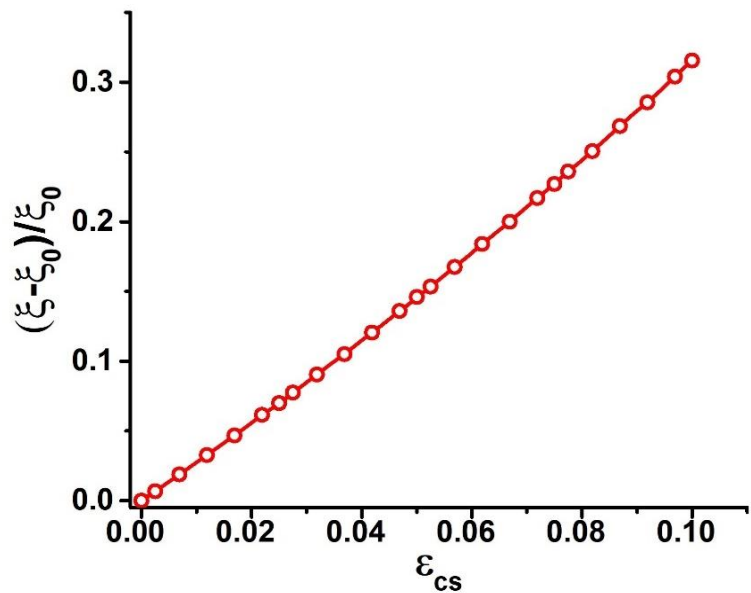

**Supplementary Figure 7** The positive correlation between the relative change of buckling amplitude and the constrained strain of the substrate ($\varepsilon_{cs}$) obtained from the FE simulations.

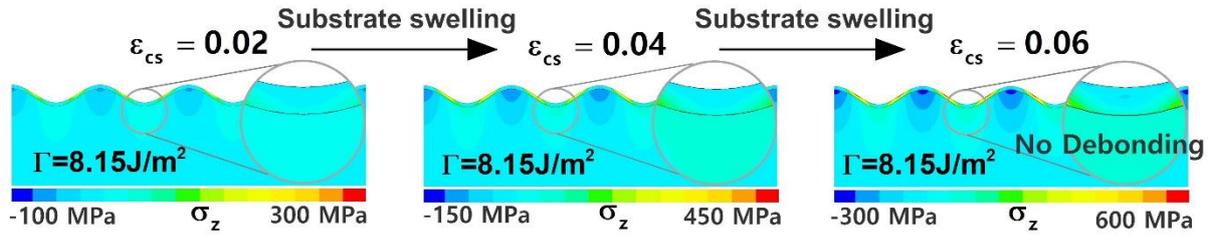

**Supplementary Figure 8** Finite Element simulation of the constrained expansion of the soft substrate with a tougher interface.

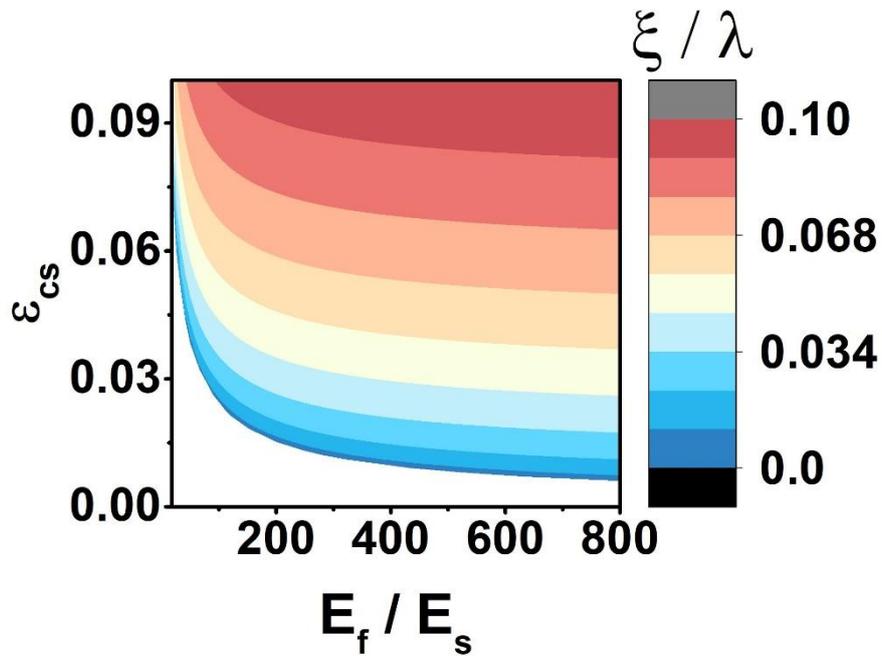

**Supplementary Figure 9** The contour plot of $\xi/\lambda$ as a function of the constrained swelling strain $\varepsilon_{cs}$ and the modulus ratio $E_f/E_s$.

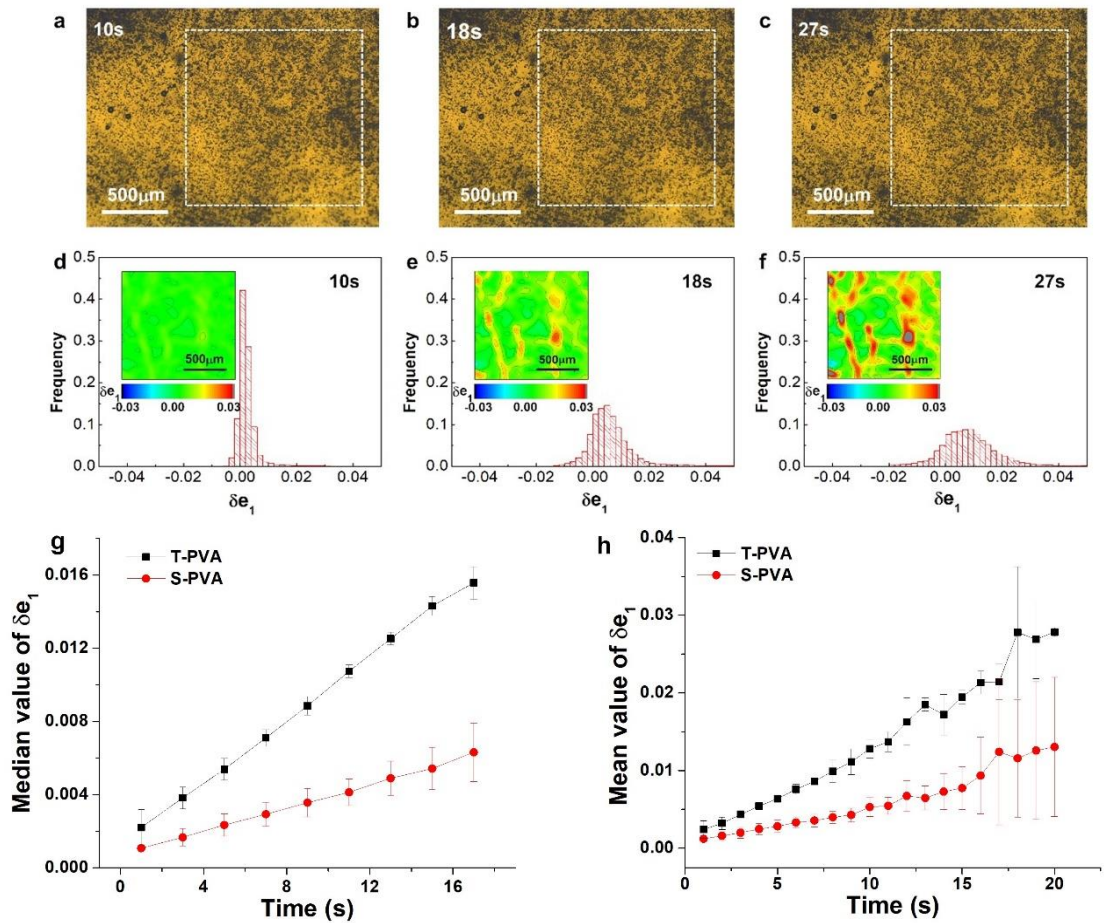

**Supplementary Figure 10 The free swelling of the hydrogel (T-PVA and S-PVA).** The optical images of a swelling PVA with surface speckles at different times, including those at (**a**)10s, (**b**)18s and (**c**) 27s. The dashed box indicates the selected area for digital image correlation analyses. The histograms of the change in the highest local principal strain $\delta e_1$ at different times, including after (**d**)10s, (**e**)18s, and (**f**) 27s. The insets in **d-f** show the spatial distribution of $\delta e_1$. (**g**) The variation of the median value of $\delta e_1$ with the swelling time. (**h**) The variation of the mean value of $\delta e_1$ with the swelling time. Note T-PVA exhibits a higher principal strain $\delta e_1$ than S-PVA

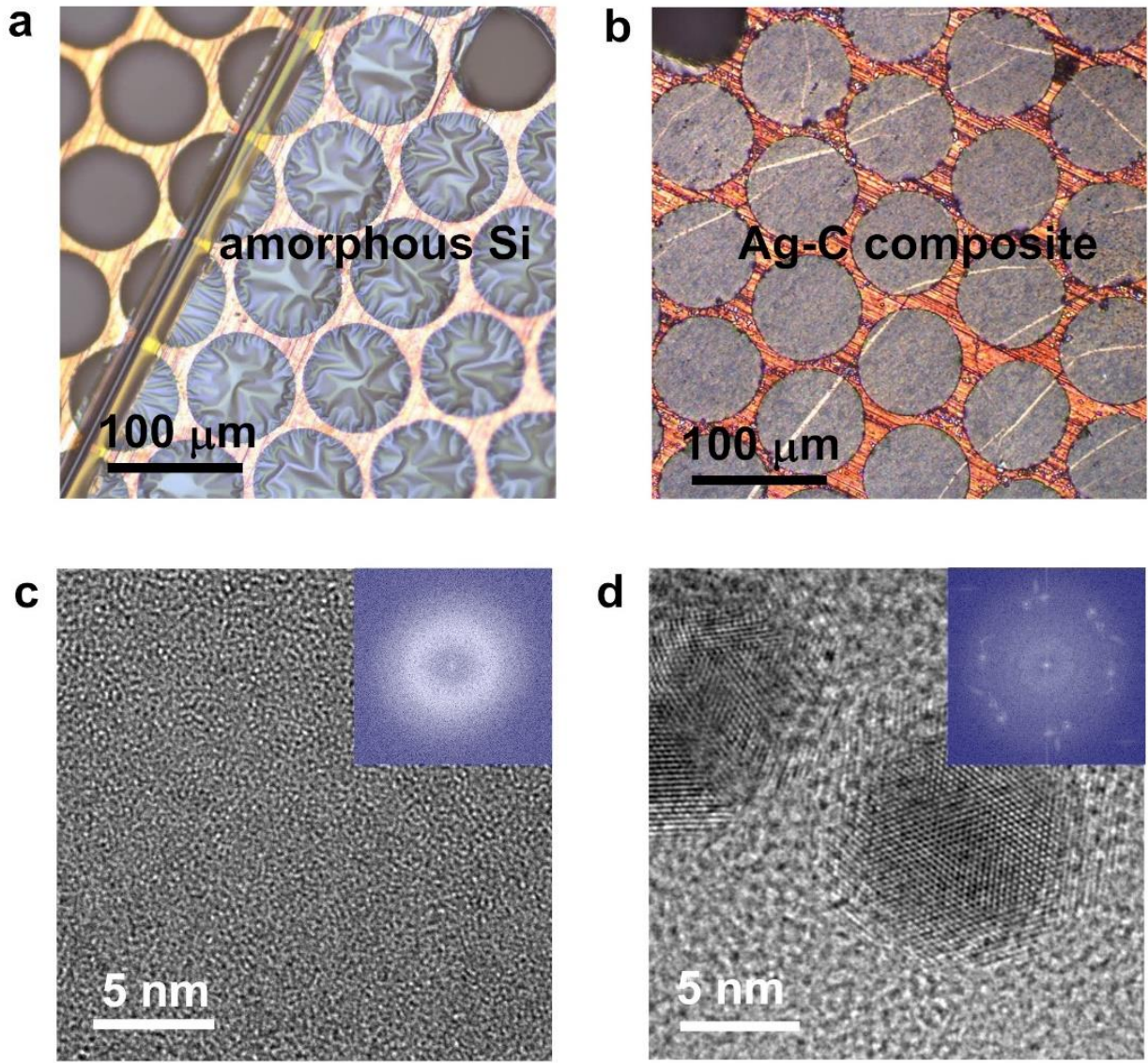

**Supplementary Figure 11 The other 2D materials fabricated using PSBEE.** (**a**) The optical image of the 2D amorphous Si on a copper grid. (**b**) The optical image of the 2D Ag-C composite on a copper grid. The HRTEM images of (**c**) the 2D amorphous Si and (**d**) the 2D Ag-C composite. Note that the insets in (**c**)-(**d**) are the corresponding FFT image.